
\let\includefigures=\iftrue
\let\useblackboard=\iftrue
\newfam\black
\input harvmac

\noblackbox

\includefigures
\message{If you do not have epsf.tex (to include figures),}
\message{change the option at the top of the tex file.}
\input epsf
\def\figin{\epsfcheck\figin}\def\figins{\epsfcheck\figins}
\def\epsfcheck{\ifx\epsfbox\UnDeFiNeD
\message{(NO epsf.tex, FIGURES WILL BE IGNORED)}
\gdef\figin##1{\vskip2in}\gdef\figins##1{\hskip.5in}
\else\message{(FIGURES WILL BE INCLUDED)}%
\gdef\figin##1{##1}\gdef\figins##1{##1}\fi}
\def\DefWarn#1{}
\def\figinsert{\goodbreak\midinsert}
\def\ifig#1#2#3{\DefWarn#1\xdef#1{fig.~\the\figno}
B
\writedef{#1\leftbracket fig.\noexpand~\the\figno}%
\figinsert\figin{\centerline{#3}}\medskip\centerline{\vbox{
\baselineskip12pt\advance\hsize by -1truein
\noindent\footnotefont{\bf Fig.~\the\figno:} #2}}
\bigskip\endinsert\global\advance\figno by1}
\else
\def\ifig#1#2#3{\xdef#1{fig.~\the\figno}
\writedef{#1\leftbracket fig.\noexpand~\the\figno}%
\global\advance\figno by1}
\fi
\useblackboard
\message{If you do not have msbm (blackboard bold) fonts,}
\message{change the option at the top of the tex file.}
\font\blackboard=msbm10 scaled \magstep1
\font\blackboards=msbm7
\font\blackboardss=msbm5
\textfont\black=\blackboard
\scriptfont\black=\blackboards
\scriptscriptfont\black=\blackboardss

\else

\fi
%
\def\yboxit#1#2{\vbox{\hrule height #1 \hbox{\vrule width #1
\vbox{#2}\vrule width #1 }\hrule height #1 }}
\def\fillbox#1{\hbox to #1{\vbox to #1{\vfil}\hfil}}
\def\ybox{{\lower 1.3pt \yboxit{0.4pt}{\fillbox{8pt}}\hskip-0.2pt}}
%
%

\def\comments#1{}



\def\II{\relax{I\kern-.10em I}}

\def\IZ{\relax\ifmmode\mathchoice
{\hbox{\cmss Z\kern-.4em Z}}{\hbox{\cmss Z\kern-.4em Z}}
{\lower.9pt\hbox{\cmsss Z\kern-.4em Z}}
{\lower1.2pt\hbox{\cmsss Z\kern-.4em Z}}
\else{\cmss Z\kern-.4emZ}\fi}
\def\IB{\relax{\rm I\kern-.18em B}}
\def\IC{{\relax\hbox{$\inbar\kern-.3em{\rm C}$}}}
\def\ID{\relax{\rm I\kern-.18em D}}
\def\IE{\relax{\rm I\kern-.18em E}}
\def\IF{\relax{\rm I\kern-.18em F}}
\def\IG{\relax\hbox{$\inbar\kern-.3em{\rm G}$}}
\def\IGa{\relax\hbox{${\rm I}\kern-.18em\Gamma$}}
\def\IH{\relax{\rm I\kern-.18em H}}
\def\II{\relax{\rm I\kern-.18em I}}
\def\IK{\relax{\rm I\kern-.18em K}}
\def\IP{\relax{\rm I\kern-.18em P}}

%

\def\inbar{\,\vrule height1.5ex width.4pt depth0pt}

\font\cmss=cmss10 
\def\IR{\relax{\rm I\kern-.18em R}}

%


%

\def\lp10{\ell_p^{10}}
\def\lp11{\ell_p^{11}}
\def\R11{R_{11}}

\def\frac#1#2{{#1 \over #2}}


\hyphenation{Di-men-sion-al}



\lref\similar{L. Girardello, M. Petrini, M. Porrati and A. Zaffaroni,
``Confinement and Condensates without Fine Tuning in Supergravity Duals
of Gauge Theories,'' JHEP {\bf 9905} (1999) 026, hep-th/9903026\semi
A. Brandhuber and K. Sfetsos, ``Non-standard compactifications with
mass gaps and Newton's law,'' JHEP {\bf 9910} (1999) 013, hep-th/9908116.}

\lref\kss{S. Kachru, M. Schulz and E. Silverstein, ``Self-tuning flat domain
walls in 5d gravity and string theory,'' hep-th/0001206.}
\lref\adks{N. Arkani-Hamed, S. Dimopoulos, N. Kaloper and R. Sundrum,
``A Small Cosmological Constant from a Large Extra Dimension,''
hep-th/0001197.}

\lref\gubser{S. Gubser, talk at Stanford 2/11/00 and to appear.}

\lref\hw{P. Horava and E. Witten, ``Heterotic and Type I String
Dynamics from Eleven-Dimensions,'' Nucl. Phys. {\bf B460} (1996) 506,
hep-th/9510209\semi
E. Witten, ``Strong Coupling Expansion of Calabi-Yau Compactification,''
Nucl. Phys. {\bf B471} (1996) 135, hep-th/9602070\semi
A. Lukas, B. Ovrut, K. Stelle and D. Waldram, ``The Universe as a
Domain Wall,'' Phys. Rev. {\bf D59} (1999) 086001, hep-th/9803235.}

\lref\led{N. Arkani-Hamed, S. Dimopoulos and G. Dvali, ``The Hierarchy
Problem and New Dimensions at a Millimeter,'' Phys. Lett. {\bf B429} (1998)
263, hep-ph/9803315\semi
I. Antoniadis, N. Arkani-Hamed, S. Dimopoulos and G. Dvali, ``New
Dimensions at a Millimeter and Superstrings at a TeV,'' Phys. Lett.
{\bf B436} (1998) 257, hep-ph/9804398.
}

\lref\tyezurab{Z. Kakushadze and H. Tye, ``Brane World,'' Nucl. Phys.
{\bf B548} (1999) 180, hep-th/9809147.}

\lref\csaki{C. Csaki, J. Erlich, T. Hollowood and Y. Shirman, ``Universal
Aspects of Gravity Localized on Thick Branes,'' hep-th/0001033.}
\lref\gremm{M. Gremm, ``Four-dimensional Gravity on a Thick Domain
Wall,'' hep-th/9912060.}
\lref\recent{M. Gremm, ``Thick Domain Walls and Singular Spaces,''
hep-th/0002040\semi
C. Gomez, B. Janssen and P. Silva, ``Dilatonic Randall-Sundrum Theory
and Renormalisation Group,'' hep-th/0002042.}

\lref\RS{L. Randall and R Sundrum, ``An Alternative to Compactification,''
Phys. Rev. Lett. {\bf 83} (1999) 4690, hep-th/9906064.}
\lref\flop{E. Witten, ``Phases of N=2 Theories in Two-Dimensions,''
Nucl. Phys. {\bf B403} (1993) 159, hep-th/9301042\semi
P. Aspinwall, B. Greene and D. Morrison, ``Calabi-Yau Moduli Space,
Mirror Manifolds and Space-time Topology Change in String Theory,''
Nucl. Phys. {\bf B416} (1994) 414, hep-th/9309097.}
\lref\conifold{A. Strominger, ``Massless Black Holes and Conifolds in
String Theory,'' Nucl. Phys. {\bf B451} (1995) 96, hep-th/9504090.}
\lref\orbifold{L. Dixon, J. Harvey, C. Vafa and E. Witten, ``Strings on
Orbifolds,'' Nucl. Phys. {\bf B261} (1985) 678.}

\lref\dfgk{O. DeWolfe, D. Freedman, S. Gubser and A. Karch,
``Modeling the fifth dimension with scalars and gravity,''
hep-th/9909134.}
\lref\cvetic{
M. Cvetic and H. Soleng, ``Supergravity Domain Walls,''
Phys. Rept. {\bf 282} (1997) 159, hep-th/9604090.}
\lref\rutgers{O. Aharony, T. Banks, A. Rajaraman and M. Rozali, unpublished
ideas.}
\lref\rubshap{V. Rubakov and M. Shaposhnikov, ``Extra Space-Time Dimensions:
Towards a Solution to the Cosmological Constant Problem," Phys. Lett.
{\bf B125} (1983) 139.}
\lref\verlinde{E. Verlinde and H. Verlinde, ``RG Flow, Gravity and
the Cosmological Constant,'' hep-th/9912018.}
\lref\schmidhuber{C. Schmidhuber, ``AdS(5) and the 4d Cosmological
Constant,'' hep-th/9912156.}
\lref\adscft{J. Maldacena, ``The Large N Limit of Superconformal
Field Theories and Supergravity,'' Adv. Theor. Math. Phys. {\bf 2} (1998)
231, hep-th/9711200\semi
S. Gubser, I. Klebanov and A. Polyakov, ``Gauge Theory Correlators
from Noncritical String Theory,'' Phys. Lett. {\bf B428} (1998) 105,
hep-th/9802109\semi
E. Witten, ``Anti-de Sitter Space and Holography,'' Adv. Theor. Math.
Phys. {\bf 2} (1998) 253, hep-th/9802150.}
\lref\ksorb{S. Kachru and E. Silverstein, ``4d Conformal Field Theories
and Strings on Orbifolds,'' Phys. Rev. Lett. {\bf 80} (1998) 4855,
hep-th/9802183.}

\lref\oldjoe{S. de Alwis,  J. Polchinski and R. Schimmrigk, ``Heterotic
Strings with Tree Level Cosmological Constant,'' Phys. Lett.
{\bf B218} (1989) 449.}
\lref\kks{S. Kachru, J. Kumar and E. Silverstein, ``Orientifolds,
RG Flows, and Closed String Tachyons,'' hep-th/9907038.}
\lref\smallinst{E. Witten, ``Small Instantons in String Theory,''
Nucl. Phys. {\bf B460} (1996) 541, hep-th/9511030.}

\lref\nati{
N. Seiberg,
``Matrix Description of M-theory on $T^5$ and $T^5/Z_2$,''
Phys.Lett. {\bf B408} (1997) 98, hep-th/9705221.}
\lref\ntwo{A. Klemm, W. Lerche, P. Mayr, C.Vafa and N. Warner,
``Self-Dual Strings and N=2 Supersymmetric Field Theory,''
Nucl.Phys. {\bf B477} (1996) 746, hep-th/9604034.}
\lref\perletc{N. Bahcall, J. Ostriker, S. Perlmutter and P. Steinhardt,
``The Cosmic Triangle: Assessing the State of the Universe,'' Science
{\bf 284} (1999) 1481, astro-ph/9906463.
}
\lref\polwitt{J. Polchinski and E. Witten,
``Evidence for Heterotic - Type I String Duality,''
Nucl.\ Phys.\  {\bf B460} (1996) 525,
hep-th/9510169.}
\lref\kaloper{N. Kaloper, ``Bent Domain Walls as Braneworlds,''
Phys. Rev. {\bf D60} (1999) 123506, hep-th/9905210.}
\lref\cohkap{A. Cohen and D. Kaplan, ``Solving the Hierarchy
Problem with Noncompact Extra Dimensions,'' Phys. Lett. {\bf B470}
(1999) 52, hep-th/9910132.}

\Title{\vbox{\baselineskip12pt\hbox{hep-th/0002121}
\hbox{SLAC-PUB-8381}\hbox{SU-ITP-00/06}}}
{\vbox{
\centerline{Bounds on curved domain walls in 5d gravity}}}
\bigskip
\centerline{Shamit Kachru, Michael Schulz and Eva Silverstein}
\bigskip
\centerline{Department of Physics and SLAC}
\smallskip
\centerline{Stanford University}
\smallskip
\centerline{Stanford, CA 94305/94309}

\bigskip
\noindent

We discuss maximally symmetric curved deformations of the flat domain
wall solutions of five-dimensional dilaton gravity that
appeared in a recent approach to the cosmological constant problem.
By analyzing the bulk field configurations and the boundary
conditions at a four-dimensional maximally symmetric curved domain wall,
we obtain constraints on such solutions.  For a special dilaton
coupling to the brane tension that appeared in recent works,
we find no curved deformations, confirming and extending
slightly a result of Arkani-Hamed et al which was argued
using a $Z_2$-symmetry of the solution.
For more general dilaton-dependent
brane tension,
we find that the curvature is bounded by
the Kaluza-Klein scale in the fifth dimension.

\Date{February 2000}

\newsec{Introduction}

There has recently been renewed interest in the old idea
that placing our world on a domain wall in a higher-dimensional
bulk space can provide a useful new perspective on the
cosmological constant problem \rubshap.
Work in this direction appeared in \refs{\verlinde,\schmidhuber}.
More recently, concrete examples of 4d domain wall universes
which have bound state gravitons \RS\ and a 4d cosmological constant
which is insensitive to quantum loops of matter fields
localized on the wall appeared in \refs{\kss,\adks}.
Similar domain wall solutions have been explored (in the context of
the AdS/CFT correspondence) in \refs{\similar,\recent}.

In this paper, we extend our work \kss\ in one
respect. There, we
concentrated for the most part on 5d gravity theories with a bulk
scalar dilaton $\phi$, and action \eqn\basicac{\eqalign{
S=&M_{*}^3\int d^5 x \sqrt{-G}\biggl[ R - {4\over 3}(\nabla
\phi)^2\biggr]\cr & + \int d^{4}x \sqrt{-g}(-f(\phi))\cr }} (with
vanishing bulk cosmological term). In \basicac, $G$ is the 5d bulk
metric while $g$ is the induced metric on the domain wall, which
is located at $x_{5} = 0$. We demonstrated that one can find flat
domain wall solutions for fairly generic thin wall delta function
sources $f(\phi)$, i.e., without ``fine-tuning'' the brane tension
$f(\phi)$. This is important because quantum loops of brane matter
fields will in the most general circumstances correct the form of
$f(\phi)$; it demonstrates some insensitivity of the existence of
a flat 4d world to brane quantum loops. However, we did not
address the issue of $\it curved$ (de Sitter or anti-de Sitter)
solutions to the same 5d equations of motion.


Here we find curved solutions with maximal symmetry in four
dimensions.  More specifically, for both negatively curved and
positively curved deformations we find that the largest scale of
curvature possible is given by the scale set by the inverse proper
length of the fifth dimension.  In particular, the curvature can
at most reach the mass scale of Kaluza-Klein modes in the fifth
dimension.  Unfortunately this upper bound is essentially
equivalent to a 4d vacuum energy of the order of the scale of
brane physics.


The organization of this paper is as follows.
In \S2\ we describe the bulk gravity solutions
with maximally symmetric curved domain walls
and the matching boundary conditions
at the domain wall.  In \S3\ we explain the
bounds on the curvature of curved solutions
that result from these solutions.
In \S4, we discuss some additional issues of interest
in analyzing the physics of the solutions discussed here
and in \refs{\kss,\adks}, including the singularities.

\newsec{Curved Solutions and Matching Conditions}

We will make the following
ansatz for the metric (following \refs{\dfgk,\kaloper}):

\eqn\metans{
ds^2= e^{2A(x_{5})}\bar g_{\mu\nu}dx^\mu dx^\nu + dx_5^2
}

\noindent
where $$\bar g_{\mu\nu}={\rm diag}(-1, e^{2\sqrt{\bar\Lambda}x_1},
e^{2\sqrt{\bar\Lambda}x_1}, e^{2\sqrt{\bar\Lambda}x_1})$$
for de Sitter space and
$$\bar g_{\mu\nu}={\rm diag}(-e^{2\sqrt{-\bar\Lambda}x_4},
e^{2\sqrt{-\bar\Lambda}x_4}, e^{2\sqrt{-\bar\Lambda}x_4}, 1)$$
for anti-de Sitter space.

Plugging this ansatz into the dilaton equations and
Einstein's equations gives
\eqn\eomI{
{8\over 3}\phi^{\prime\prime}+{32\over 3}A^\prime\phi^\prime
-{{\partial f}\over{\partial\phi}}\delta(x_5) = 0
}
\eqn\eomII{
6(A^\prime)^2-{2\over 3}(\phi^\prime)^2
-6\bar\Lambda e^{-2A} = 0
}
\eqn\eomIII{
3A^{\prime\prime}+{4\over 3}(\phi^\prime)^2
+3\bar\Lambda e^{-2A} + {1\over 2} f(\phi) \delta(x_5) = 0
}
Note here that the zero mode $A(0)$ always appears together
with $\bar\Lambda$ here; we will take $A(0)=0$ in what
follows.

Integrating the first equation in the bulk gives

\eqn\solphi{
\phi^\prime=\gamma e^{-4A}
}
for some integration constant $\gamma$.
Substituting this into the second equation gives
\eqn\Aprime{
A^\prime=\epsilon\sqrt{{1\over 9}\gamma^2 e^{-8A}
+ \bar\Lambda e^{-2A}}
}
where $\epsilon=\pm 1$ determines the branch of the
square root that we choose in the solution.
Note here that this solution only makes sense
when the argument of the square root in
\Aprime\ is positive; for anti-de Sitter
slices (negative $\bar\Lambda$) this gives
a constraint on $\bar\Lambda$ which we will
discuss in \S3.

This equation can be integrated to yield
\eqn\relint{
\int^A{{\epsilon ~dA}\over{\sqrt{{1\over 9}\gamma^2 e^{-8A}
+\bar\Lambda e^{-2A}}}}=x_5+{3\over 4}c
}
The left-hand side of \relint\ is
\eqn\ansint{
{3\over 4}\epsilon {1\over{|\gamma|}}e^{4A}
~{_2F_1}({1\over 2},{2\over 3}, {5\over 3},
-{{9 \bar\Lambda}\over\gamma^2}e^{6A})
=x_5+{3\over 4}c
}
where $_2F_1({1\over 2},{2\over 3}, {5\over 3},
z)\equiv F(z)$ is a hypergeometric function.
It is analytic on $\IC-\{[1,\infty)\subset\IR\}$
and increases monotonically from $F(-\infty)=0$
through $F(0)=1$ until it attains its maximum at
$F(1)= F_{max} = {{\Gamma({5\over 3})\Gamma({1\over 2})}
\over\Gamma({7\over 6})}=1.725$.

Because $F\ge 0$, the solution \ansint\ is valid only
on one side of $x_5=-{3\over 4}c$ (determined by the sign $\epsilon$).
At $x_5=-c$ there is a curvature singularity.  As
in \refs{\kss,\adks}, we make the assumption
that the space can
be truncated at this singularity, at least as far
as low-energy physics is concerned.

Let us now introduce a domain wall at $x_5=0$.
We must match the bulk solutions (given implicitly
in \solphi\ and \ansint) on the two sides of the
wall, consistent with the $\delta$-function
terms in \eomI\eomIII.  Let us denote the
integration constants on the left ($x_5<0$) side
of the wall by $c_1,\gamma_1,d_1$ and those on
the right ($x_5>0$) side by $c_2,\gamma_2,d_2$.
Here $d_i$ refers to the zero mode $\phi(0)$ of
the dilaton field on the $i$th side of the wall.
Imposing continuity of $\phi$ at the wall
fixes $d_2$.

Defining $\tilde c_i=c_i/F|_{x_5=0}$, we find
the matching conditions
\eqn\matchI{
-{8\over 3}M_*^3\biggl({1\over{\tilde c_1}}+{1\over{\tilde c_2}}
\biggr)={{\partial f}\over{\partial\phi}}(\phi(0))
}
\eqn\matchII{
M_*^{3} \biggl(
\sqrt{({1\over{\tilde c_1}})^2+9\bar\Lambda}+
\sqrt{({1\over{\tilde c_2}})^2+9\bar\Lambda}\biggr)
={1\over 2}f(\phi(0))
}
We here used the fact, which follows from
\ansint\ evaluated at $x_5=0$ with
$A(0)=0$, that
$|\gamma_i||\tilde c_i|=1$.

\newsec{Bounds on Curved Deformations}

\subsec{Asymmetric Solutions (I) (General $f(\phi)$)}

We have now gathered the information we need to
determine the extent of curvature of these curved-slice
deformations of the flat solutions of \refs{\kss,\adks}.
We will first discuss a bound on $|\bar\Lambda|$,
which basically constrains it to be less than
the inverse proper length of the fifth dimension,
that applies to both signs of $\bar\Lambda$.
We will then discuss a tighter bound that
arises in the case of positive $\bar\Lambda$.

Consider the equation \ansint\ at $x_5=0$:

\eqn\anszero{
\epsilon |\tilde c| F(-9\Lambda (\tilde c)^2)
=c
}
Defining $y\equiv \sqrt{9|\bar\Lambda|}|\tilde c|$, this
equation implies
\eqn\ansy{
|y F(-y^2)|=|c|\sqrt{9|\bar\Lambda|}
}
Now the quantity $yF(-y^2)$ is bounded.  In fact, its
maximum value (attained as $y \to \infty$) is 4.
We have from \ansy\ that
\eqn\boundI{
\sqrt{9|\bar\Lambda|} < 4 |{1\over c_i}|
}
where we added the index to $c_i$ since this
bound applies on either side of the domain wall.

Note from the metric \metans\ that
$|c_i|$ is the proper distance to the singularity
on the $i$th side of the wall.
So for either sign of $\bar\Lambda$, we find that
the effective 4d cosmological
constant of the curved solutions
is bounded to be smaller than the Kaluza Klein scale in the bulk.

This reflects the same physical point made in \refs{\kss,\adks}:
there is no contribution from physics localized on the brane to
the 4d cosmological constant.  A brane-scale cosmological
constant would have manifested itself in
a contribution to \boundI\ which depends on $f(\phi(0))$, and
such terms are absent.
These bounds arise from the matching conditions, but note that it
is not
the case that the singularities recede to $\infty$ (or come in
to the origin) as one saturates the bound.

The largest phenomenologically viable value for the proper
distance $c$ is roughly a millimeter \led. This would give us a
bound on $\bar\Lambda$ of about $10^{-6}{\rm eV}^2$. This is much
larger than the observed value $\bar\Lambda\sim 10^{-64} {\rm
eV}^2$ of the cosmological constant. Note that we here are using
``general relativity'' conventions for the cosmological constant
$\bar\Lambda$; the standard ``particle physics'' cosmological
constant is $\Lambda_4=M_4^2\bar\Lambda\sim {\rm mm}^{-4}$.
Unfortunately this is within a couple of orders of magnitude of
the Standard Model scale of TeV$^4$.  In a model with
supersymmetry spontaneously broken at the TeV scale, this would be
the scale of a brane cosmological constant.

For positive $\bar\Lambda$ this $1/c$ scale is itself
bounded by a further constraint.  Consider the matching
condition \matchII.  It implies that
\eqn\posbound{
{1\over {\tilde c_i^2}} < {1\over {2M_*^3}}f(\phi)
}
Therefore, since $c=\tilde c/F|_0$, we can extend
\boundI\ to
\eqn\boundII{
\sqrt{9\bar\Lambda} < 4 F_{max} {1\over {2M_*^3}}f(\phi)
}

In fact we can do better than \boundII. The 4d Planck scale $M_4$
is given by: \eqn\planck{M_{4}^2 ~= M_{*}^3~\int dx_{5}~ e^{2A}
~=~ {M_{*}^{3}\over 9\bar\Lambda} \biggl( \sqrt{{1\over \vert
\tilde c_{1}\vert^2} + 9 \bar \Lambda} + \sqrt{{1\over \vert
\tilde c_{2}\vert^2} + 9 \bar \Lambda} - {1\over \vert \tilde
c_{1}\vert} - {1\over\vert\tilde c_{2}\vert} \biggr)}\ Multiplying
\planck\ by $\bar \Lambda$, and dividing by $M_{4}^2$, we get an
equation for $\bar \Lambda$.  For positive $\bar \Lambda$, we can
use the matching condition \matchII\ to replace the first two
terms in the parentheses in \planck\ with ${1\over 2}f(\phi(0))$.
We then obtain the inequality \eqn\posbound{\bar \Lambda < {1\over
18} {f(\phi(0))\over M_{4}^2}} (for negative $\bar \Lambda$, we
would not obtain such an inequality). So for instance if the value
of $f(\phi(0))$ is TeV scale, which is natural if we take the
standard model (cut off at about a TeV) to live on the brane, then
\eqn\posboundII{\bar \Lambda < 10^{-33} ({\rm TeV})^2} This is of
the same order as the contribution of a brane with supersymmetry
spontaneously broken at a TeV.

\subsec{Symmetric Solutions (II) ($f(\phi)=e^{\pm{4\over 3}\phi}$)}

When we pick $f(\phi)=e^{\pm{4\over 3}\phi}$, we find the matching
condition \matchI\ becomes

\eqn\symmmatchI{
M_*^3\biggl({1\over{|\tilde c_1|}}+{1\over{|\tilde c_2|}}\biggr)
=\pm{1\over 2}e^{\pm{4\over 3}\phi}
}
which agrees with the second condition \matchII\ when
$\bar\Lambda=0$.  When $\bar\Lambda\ne 0$, the two conditions
\symmmatchI\ and \matchII\ contradict each other, and there
are no solutions.  This means that the symmetric solutions
of \refs{\kss,\adks}
(solutions (II) in the classification of \kss) do not
have any deformations with 4d
de Sitter or anti-de Sitter symmetry.  This slightly extends
the result of Arkani-Hamed et al \adks, who observed
that such deformations
would violate the $Z_2$ symmetry of the solution,
and thus could not appear in a $Z_2$ orbifold of
this solution.

\newsec{Discussion and Further Issues}

A priori there is a question as to whether the space of
integration constants is parameterized by vacuum expectation
values of fluctuating fields in four dimensions, or whether
instead different members of this family arise from different
four-dimensional Lagrangians.\foot{We thank
S. Dimopoulos and R. Sundrum for discussions on this point.}
The existence of anti-de Sitter
and de Sitter deformations (bounded though they are) suggests
that these deformations constitute parameters in the 4d
effective theory.  If the effective 4d cosmological constant
were parameterized by a field, then in solving its equations
of motion one would end up with one consistent possibility for
the value of the 4d cosmological constant.  The fact that
we find a family of solutions suggests that this is not
the case here.  Indeed, naive calculation of the coefficient of
the kinetic term for the mode which moves one from flat to curved
4d metrics does suggest that it is not a dynamical mode (it has
infinite kinetic term).  However the divergence in
the calculation arises at the
singularities, so this conclusion depends sensitively
on how the singularities are resolved by microphysics.

To a 4d effective field theorist, the choice of which member of
the family to start with constitutes a tuning of the 4d
cosmological constant.  From the point of view of the microscopic
5d theory, this tuning involves a parameter in the solution and
not a parameter in the Lagrangian.  If this system can be embedded
consistently into string theory, where there are no input
parameters in the ``Lagrangian,'' the mere existence of Poincare
invariant solutions after some quantum corrections have been taken
into account would be significant, even if such solutions lie in a
family of curved solutions that signal the appearance of
fine-tuning at low energies. In any case, our results here
indicate that the apparent fine-tuning required to choose a flat
slice is independent of Standard Model physics, though it can
arise at the same scale.

Having understood better the situation with respect to
this issue of fine-tuning, one is led to consider the
main challenge identified in \refs{\kss,\adks}:  the question
of possible microphysical constraints on the (codimension one)
singularities
in the solutions.  The type of analysis we did here
might help resolve an issue raised in \gubser,
as we will mention presently, after first discussing
the issue in a little more generality.

One possibility is that boundary conditions are required
at the singularities, as in the case analyzed in
\cohkap.  It is then important to check whether
the appropriate boundary conditions, along
with the equations of motion and matching
conditions, can be solved
within the space of curved solutions we have
identified \refs{\kss,\adks}.

There are some singularities in string
theory (like conifolds, orbifolds, and brane-orientifold systems) which
have a well-understood quantum resolution involving
new degrees of freedom at the singularity;
in these cases the resolution does
not imply any extra boundary conditions in
the effective long-wavelength theory.

It has recently been suggested that the singularities
that appear in our solutions do not permit a finite-temperature
deformation accessible with a long-wavelength general
relativistic analysis \gubser.  This is a criterion
that does not appear to contradict the
microscopic consistency of orbifolds or conifolds,
and the case of orientifolds and their duals
must be considered carefully.
Because of the large curvatures (and in some cases large
couplings) in the backgrounds we consider here, such
an analysis is necessarily limited.  However, the
general question of how finite temperature can be
obtained in these backgrounds is an important one.

Within the context of the
analysis of \gubser, it is notable that our solutions lie on the
boundary between (conjecturally) allowed and (conjecturally)
disallowed singularities.  It is important to redo this
analysis for solutions which include some bulk corrections.
In particular, a nontrivial bulk dilaton potential of the
right sign (as in our case (III) \kss) may put us in
the allowed region according to the conjectured criterion
of \gubser.  Instead of fine-tuning to obtain 4d Poincare
invariant slices as we did in case (III) of \kss, one
can consider curved solutions of the sort given
here.  In the context of the type (III) situation
where there is a bulk potential for $\phi$, this
is in fact natural if we do not wish to fine-tune
the parameters in the 5d Lagrangian in order to obtain
a 4d Poincare invariant solution.  It is possible
that this bulk correction will induce a sub-TeV correction
to the 4d cosmological constant, while satisfying
the conjectured constraints coming from
the long-wavelength analysis of \gubser.

\centerline{\bf Acknowledgements}

We would like to thank N. Arkani-Hamed, T. Banks, S. Dimopoulos,
S. Gubser, N. Kaloper, D. Kaplan, H. Ooguri, J. Polchinski, S.
Shenker, R. Sundrum, and L. Susskind for helpful discussions
and/or correspondence on this subject. We would like to thank E.
Witten for alerting us to an important error in the first version
of this paper.  S. K. is supported in part by an A.P. Sloan
Foundation Fellowship, M.S. is supported in part by an NSF
Graduate Research Fellowship, and E.S. is supported in part by a
DOE OJI Award and an A.P. Sloan Foundation Fellowship. S.K. and
E.S. are both supported in part by the DOE under contract
DE-AC03-76SF00515.

\listrefs
\end